\documentclass[10pt,twocolumn,twoside]{IEEEtran} %2칼럼(single-spaced double-column)

\usepackage{amsmath}
\usepackage[dvips]{epsfig}
\usepackage{graphicx}

\begin{document}

\title{Oversampling Effect Analysis of Correlation Metric in Selected Mapping Scheme with Nonlinearity}

%\author{Jun-Young Woo, Kee-Hoon Kim, Jong-Seon No, {\em Fellow, IEEE}, and Dong-Joon Shin, {\em Senior Member, IEEE}
\author{Author 1, Author 2, Author 3, and Author 4
%\thanks{J.-Y. Woo, K.-H. Kim, and J.-S. No are with the Department of Electrical and Computer Engineering, INMC, Seoul National University, Seoul,  Korea (e-mail: \{jywoo, kkh\}@ccl.snu.ac.kr, jsno@snu.ac.kr).}
%\thanks{D.-J.~Shin is with the Department of Electronic Engineering, Hanyang University, Seoul, Korea (e-mail: djshin@hanyang.ac.kr).}
}

\markboth{IEEE COMMUNICATIONS LETTERS, VOL. , NO. , 2014}{Woo {\em \lowercase{et al.}}: Oversampling Effect Analysis of Correlation Metric in SLM scheme}

\maketitle

\begin{abstract}
In this letter, oversampling effect is analyzed when correlation (CORR) metric is used in the selected mapping (SLM) scheme with the presence of nonlineartiy.
 In general, 4 times oversampling is enough for estimating continuous signal. But, we can use 2 times oversampling with similar bit error rate (BER) performance.
Therefore, we can reduce the computational complexity half roughly.
 Simulation results show that BER performance when 2 times oversampling is used is almost the same as 4 times oversampling is used. On the other hand, BER performance when 1 times ovesampling (Nyquist rate) is used is degraded than 2 times oversampling or 4 times oversampling are used. By deriving the Pearson correlation coefficient, simulation results can be confirmed.
\end{abstract}


\section{Introduction}

\IEEEPARstart{F}{or} many years, orthogonal frequency division multiplexing (OFDM) has received enormous attention. Due to the fact that OFDM has high peak-to-average power ratio (PAPR), many papers tried to mitigate the PAPR problem. Partial transmit sequence (PTS) and selected mapping (SLM) can be a good answer to solve the PAPR problem \cite{muller97}.  PTS and SLM generates alternative OFDM signal sequences. Among them, the one with the minimum PAPR is selected for transmission. Since these schemes have high computational complexity, low complexity schemes are proposed. Besides, many other techniques such as clipping, tone reservation, and tone injection are also proposed.

Above schemes are focused on reducing the PAPR, but recently, many other metrics are proposed to improve the bit error rate (BER) performance when the presence of nonlinear high power amplifier (HPA).
Intermodulation distortion (IMD) \cite{Rodrigues06}, distortion-to-signal power ratio (DSR) \cite{Dalakta12_VT}, mean squared error (MSE) \cite{Park07}, and single point correlation (CORR) \cite{Dalakta12} outperform the PAPR metric in terms of BER performance.

In Section \ref{sec:CORR}, we briefly review the CORR metric and show that CORR is the optimum metric in terms of BER performance. In Section \ref{sec:OS}, oversamplig effect of CORR is analyzed
 when the presence of nonlinear HPA in SLM scheme. Simulation result is given in Section \ref{sec:Simul} and Section \ref{sec:Con} concludes this letter.

\section{Overview of SLM with CORR Metric}\label{sec:CORR}

Binary data sequences are modulated by $M$-ary quadrature amplitude modulation (QAM) constellation. Then, an oversampled input symbol sequence ${\mathbf X}=[X_0$, $X_1,{\cdots},X_{N-1}, \underbrace{0,{\cdots},0}_{(L-1)N}]$ is inverse Fourier transformed (IFFTed), where $N$ is the number of subcarriers. The $n$th time domain sample of OFDM signal sequence can be expressed as
\begin{equation}
  x_n = \frac{1}{\sqrt{N}}\sum_{k=0}^{N-1}X_ke^{j\frac{2\pi kn}{LN}}
\end{equation}
where $LN$ is the IFFT size, $L$ is the oversampling factor, and $n=0,1,{\cdots},LN-1$.

\begin{figure}[t]
  \centering
  \includegraphics[width=\linewidth]{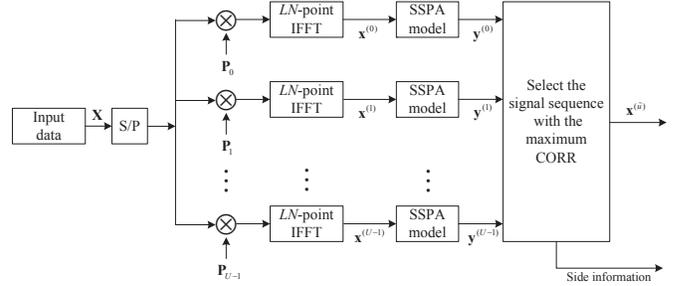}\\
  \caption{A block diagram of SLM scheme with CORR metric.}\label{fig:SLM_CORR}
\end{figure}

Fig. \ref{fig:SLM_CORR} shows a block diagram of SLM scheme with CORR metric. Conventional SLM scheme generates $U$ alternative OFDM signal sequences by componentwise multiplication of input symbol sequence and $U$ different phase sequences.
Let ${\mathbf P}^{(u)}=\{P_0^{(u)},P_1^{(u)},{\cdots},P_{N-1}^{(u)}\}$ be the $u$th phase sequence and the elements are $P_k^{(u)}=e^{j\phi_k^{(u)}}$ where $\phi_k^{(u)}\in[0,2\pi)$, $k=0,1,{\cdots},N-1$, and $u=0,1,{\cdots},U-1$. For implementation, we use $P_k^{(u)}=\pm1$.

Each alternative signal sequence is $LN$-point IFFTed ${\mathbf x}^{(u)}=\{x_0^{(u)},x_1^{(u)},{\cdots},x_{LN-1}^{(u)}\}$ and passes through the solid state power amplifier (SSPA) model. Here, the polynomial model is used for SSPA model. The HPA polynomial model is expressed as a third order nonlinearity. Thus, $n$th element of the output of the HPA can be represented by
\begin{equation}\label{eq:poly}
  y_n\approx \alpha_1x_n+\alpha_3x_n|x_n|^2
\end{equation}
where $\alpha_1=1$ and $\alpha_3=-0.1769$ are the polynomial coefficients \cite{Dalakta12}.
Then, CORR calculate $U$ single point correlation between input ${\mathbf x}^{(u)}$ and output ${\mathbf y}^{(u)}$ of the SSPA model. The CORR computation is given as
\begin{equation}\label{eq:CORR}
  R_{xy}^{(u)}=\sum_{n=0}^{LN-1}x_n^{(u)}y_n^{*(u)}=\alpha_1\sum_{n=0}^{LN-1}|x_n^{(u)}|^2 +\alpha_3\sum_{n=0}^{LN-1}|x_n^{(u)}|^4
\end{equation}
where $*$ indicates complex conjugation.
Among $U$ $R_{xy}^{(u)}$, the signal with maximum CORR is selected for transmission. PTS scheme with CORR is treated in \cite{Dalakta12}, but in this letter, we treat SLM scheme with CORR since SLM scheme is a general scheme.

In \cite{Bae13}, analytic BER of SLM scheme is described when nonlinear HPA is presented.
\begin{equation}\label{eq:BER_SLM}
  P_b = \frac{4}{\log_2M}\left(1-\frac{1}{\sqrt M}\right)Q\left(\sqrt{\frac{3}{{M-1}}{\rm SDNR}}\right).
\end{equation}

According to \cite{Bae13}, for better BER peformance, SDNR should be large. To do so, $E[x_n^{(u)}y_n^{*(u)}]$ should be large and it is identical to CORR metric. Thus, CORR metric is the optimal in terms of BER performance. Thus, we focus on CORR metric among many metrics targeting the BER performance and describe the oversampling effect in SLM scheme with nonlinearity.

\section{Oversampling Effect}\label{sec:OS}
Based on \cite{Wang10}, oversampled signal can be achieved by linear combination of Nyquist rate samples.
The impulse response of interpolator for the $L$ times oversampling is defined as
\begin{equation}\label{eq:interpolator}
h_L[n]=\frac{\sin(\pi n/L)}{\pi n/L}
\end{equation}
where we use the limited filter length ($I$) in practice.
Then, the output of the interpolator can be expressed as
\begin{equation}\label{eq:outofinterpolator}
  \tilde{x}_L[n_L]=\sum_{k=\lceil (n_L-LI)/L\rceil}^{\lfloor(n_L+LI)/L\rfloor}x_1[k]h_L[n_L-Lk]
\end{equation}
where $x_L[n_L]$ is $n_L$th element of $L$ times oversampled signal ${\mathbf x}$. In this letter, we use $I=2$ for analyzation.

Let us consider 16 times oversampled signal because we can regard it as a continuous signal. Then, the interpolation matrix ${\mathbf M}=[M_1, M_2, {\cdots}, M_{15}]$ can be written as \eqref{eq:filter_coefficient} at the top of the page 3.
\begin{figure*}[t]
  \centering
  \begin{equation}\label{eq:filter_coefficient}
   \left(\begin{smallmatrix}
    h_{16}[17]&h_{16}[18]&h_{16}[19]&h_{16}[20]&h_{16}[21]&h_{16}[22]&h_{16}[23]&h_{16}[24]&h_{16}[25]&h_{16}[26]&h_{16}[27]&h_{16}[28]&h_{16}[29]&h_{16}[30]&h_{16}[31]\\
    h_{16}[1]&h_{16}[2]&h_{16}[3]&h_{16}[4]&h_{16}[5]&h_{16}[6]&h_{16}[7]&h_{16}[8]&h_{16}[9]&h_{16}[10]&h_{16}[11]&h_{16}[12]&h_{16}[13]&h_{16}[14]&h_{16}[15]\\
    h_{16}[-15]&h_{16}[-14]&h_{16}[-13]&h_{16}[-12]&h_{16}[-11]&h_{16}[-10]&h_{16}[-9]&h_{16}[-8]&h_{16}[-7]&h_{16}[-6]&h_{16}[-5]&h_{16}[-4]&h_{16}[-3]&h_{16}[-2]&h_{16}[-1]\\
    h_{16}[-31]&h_{16}[-30]&h_{16}[-29]&h_{16}[-28]&h_{16}[-27]&h_{16}[-26]&h_{16}[-25]&h_{16}[-24]&h_{16}[-23]&h_{16}[-22]&h_{16}[-21]&h_{16}[-20]&h_{16}[-19]&h_{16}[-18]&h_{16}[-17]
\end{smallmatrix}\right).
  \end{equation}
  \vspace{-35pt}
  \begin{center}
\line(1,0){520}
\end{center}
\end{figure*}
For example, $\tilde{x}_{16}[16m+1]=h_{16}[17]x_1[16(m-1)]+ h_{16}[1]x_1[16m]+ h_{16}[-15]x_1[16(m+1)]+ h_{16}[-31]x_1[16(m+2)]$, where $0\leq m\leq N-1$.

We can only express $16$ samples among $16N$ samples without loss of generality.
 Then, we can rewrite \eqref{eq:CORR} as
%\begin{align}\label{eq:CORR_16}
%   \bar{R}_{\tilde{x}\tilde{y}}^{(u)}&=\sum_{s=0}^{15}\tilde{x}_{16m+s}^{(u)}\tilde{y}_{16n+s}^{*(u)} \nonumber\\
%  &=
%\end{align}
\begin{equation}\label{eq:CORR_16}
  \bar{R}_{\tilde{x}\tilde{y}}^{(u)}=\sum_{s=0}^{15}\tilde{x}[{16m+s}]^{(u)}\tilde{y}[{16m+s}]^{*(u)}.
\end{equation}
There are 16 terms and we assign alphabets, (a) to (p), to each terms, i.e., ${\rm (a)}=\tilde{x}[{16m+1}]^{(u)}\tilde{y}[{16m+1}]^{*(u)}$ and ${\rm (p)}=\tilde{x}[{16m+15}]^{(u)}\tilde{y}[{16m+15}]^{*(u)}$. For better understanding, let us consider only (i) term and write it concretely as \eqref{eq:I_term} by \eqref{eq:outofinterpolator} and \eqref{eq:filter_coefficient}.
\begin{figure*}[t]
  \centering
  \begin{align}\label{eq:I_term}
  I &= \tilde{x}[{16m+8}]^{(u)}\tilde{y}[{16m+8}]^{*(u)}\nonumber\\
    &=(-0.21x[16m-16]+0.63x[16m]+0.63x[16m+16]-0.21x[16+32])\nonumber\\
    &~~~~(-0.21y[16m-16]+0.63y[16m]+0.63y[16m+16]-0.21y[16+32])^*\nonumber\\
    &=0.04x[16m-16]y^*[16m-16] -0.13x[16m-16]y^*[16m] -0.13x[16m-16]y^*[16m+16] +0.04x[16m-16]y^*[16m+32]\nonumber\\
    &-0.13x[16m]y^*[16m-16] ~\qquad+0.39x[16m]y^*[16m] ~~~~~~+0.39x[16m]y^*[16m+16] ~~~~~~-0.13x[16m]y^*[16m+32]\nonumber\\
    &-0.13x[16m+16]y^*[16m-16] +0.39x[16m+16]y^*[16m] +0.39x[16m+16]y^*[16m+16] -0.13x[16m+16]y^*[16m+32]\nonumber\\
    &+0.04x[16m+32]y^*[16m-16] -0.13x[16m+32]y^*[16m] -0.13x[16m+32]y^*[16m+16] +0.04x[16m+32]y^*[16m+32].
    \end{align}
  \vspace{-35pt}
  \begin{center}
\line(1,0){520}
\end{center}
\end{figure*}
With same way, we can compute all the other terms.

Now, let us consider $L$ times oversampling for computing CORR metric. We only need (a) term for Nyquist rate sampling. ${\rm (a)+(i), (a)+(e)+(i)+(m)},$ and ${\rm (a)+(b)+{\cdots}+(p)}$ terms are needed to compute \eqref{eq:CORR} for 2 times, 4 times, and 16 times oversampling, respectively.

\begin{table*}[t]
\centering
\caption{Coefficients set of the equations to compute CORR metric when $L$ times oversampling is used}
\label{table:CoefOfEq}
\begin{tabular}{|c|c|c|c|c|c|c|c|c|c|c|c|c|c|c|c|c|}
  \hline
  Index & $1$ & $2$ & $3$ & $4$ & $5$ & $6$ & $7$ & $8$ & $9$ & $10$ & $11$ & $12$ & $13$ & $14$ & $15$ & $16$ \\\hline\hline
  $L=1$ & $0$ & $0$ & $0$ & $0$ & $0$ & $1$ & $0$ & $0$ & $0$ & $0$ & $0$ & $0$ & $0$ & $0$ & $0$ & $0$ \\
  $L=2$ & $0.04$ & $-0.13$ & $-0.13$ & $0.04$ & $-0.13$ & $1.4$ & $0.4$ & $-0.13$ & $-0.13$ & $0.4$ & $0.4$ & $-0.13$ & $0.04$ & $-0.13$ & $-0.13$ & $0.04$ \\
  $L=4$ & $0.09$ & $-0.33$ & $-0.30$ & $0.09$ & $-0.33$ & $2.3$ & $0.94$ & $-0.3$ & $-0.3$ & $0.94$ & $1.3$ & $-0.33$ & $0.09$ & $-0.3$ & $-0.33$ & $0.09$ \\
  $L=16$& $0.37$ & $-1.42$ & $-1.25$ & $0.36$ & $-1.42$ & $7.72$ & $3.94$ & $-1.25$ & $-1.25$ & $3.94$ & $6.72$ & $-1.42$ & $0.36$ & $-1.25$ & $-1.42$ & $0.37$\\
  \hline
\end{tabular}
\end{table*}

Table \ref{table:CoefOfEq} shows the coefficients of the equations to compute CORR metric when $L$ times oversampling is used. Indices are arranged in numerical order ($1$--$16$). For example, in \eqref{eq:I_term}, the coefficients of index $1$, index $3$, and index $6$ are $0.04, -0.13$, and $0.39$, respectively.
To compute the correlation between sequences derived from \eqref{eq:CORR_16} with different $L=\{1,2,4,16\}$, we only consider the coefficients of each equation. 

Now, we use the Pearson correlation coefficient ($r$). $r$ is a measure of linear correlation between two sequences (variables) and is giving a value $[-1,1]$, where $0$ is no correlation and $1({\rm or}-1)$ is total positive(or negative) correlation.

Pearson correlation coefficient between two sequences $(\mathbf{x,y})$ is defined as
\begin{equation}\label{eq:PearsonCoef}
  r=\frac{\sum_i(x_i-\bar{\mathbf{x}})(y_i-\bar{\mathbf{y}})}{\sqrt{\sum_i(x_i-\bar{\mathbf{x}})^2\sum_i(y_i-\bar{\mathbf{y}})^2}}
\end{equation}
where $\bar{x}$ denotes the sample mean and $x_i$ denotes the $i$th element of the sequence $\mathbf{x}$.

\begin{table}[t]
\caption{Pearson correlation coefficient between different $L$}
\centering
\begin{tabular}{|c|c|c|c|}
  \hline
   & $L=1$ \& $L=16$ & $L=2$ \& $L=16$ & $L=4$ \& $L=16$ \\\hline
  $r$ & 0.6023 & 0.9133 & 0.9839 \\
  \hline
\end{tabular}
\label{table:Correlation coefficient}
\end{table}

Table~\ref{table:Correlation coefficient} shows the Pearson correlation coefficients between two sequences obtained from Table \ref{table:CoefOfEq} when different $L$ is used. We can regard when $L=16$ is used as a continuous signal which can estimate the nonlinear effect of HPA well. When comparing $L=1$ and $L=16$, $r$ is $0.6023$, which has very low correlation. On the other hand, the value of $r$ when comparing $L=2$ \& $L=16$ and $L=4$ \& $L=16$ are $0.9133$ and $0.9839$, respectively, which have very high correlation. Therefore, we can expect that $2$ times and $4$ times oversampling can be used instead of using $16$ times oversampling.

\section{Simulation Results}\label{sec:Simul}

\begin{figure}[t]
  \centering
  \includegraphics[width=\linewidth]{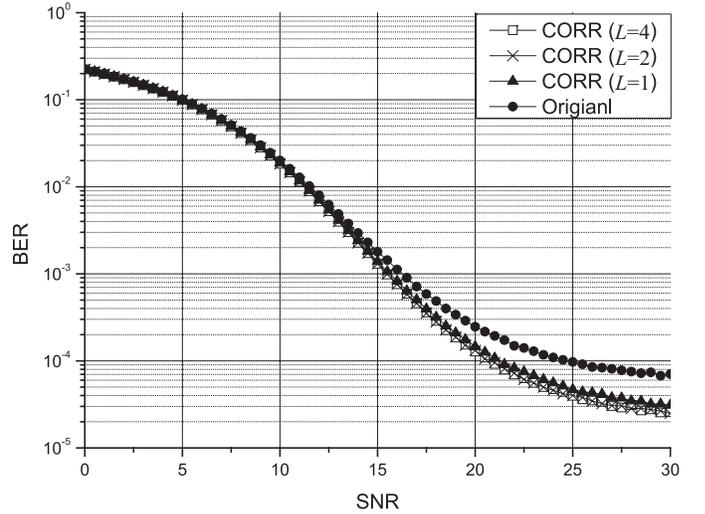}\\
  \caption{BER performances of SLM schemes with CORR metric when $N=1024$, $U=4$, and $L=1,~2,$ and $4$.}\label{fig:BER}
\end{figure}

In this section, we derive the BER performance of SLM schemes with CORR metric when different $L$ times oversampling are used. Also, nonlinear HPA is considered. There are four BER curves in Fig. \ref{fig:BER}. Original indicates OFDM system without SLM scheme.

\section{Conclusion}\label{sec:Con}

\end{document}